\documentclass[12pt,journal,peerreview]{IEEEtran}

\makeatletter
\long\def\@makecaption#1#2{\ifx\@captype\@IEEEtablestring%
	\footnotesize\begin{center}{\normalfont\footnotesize #1}\\
		{\normalfont\footnotesize\scshape #2}\end{center}%
	\@IEEEtablecaptionsepspace
	\else
	\@IEEEfigurecaptionsepspace
	\setbox\@tempboxa\hbox{\normalfont\footnotesize {#1.}~~ #2}%
	\ifdim \wd\@tempboxa >\hsize%
	\setbox\@tempboxa\hbox{\normalfont\footnotesize {#1.}~~ }%
	\parbox[t]{\hsize}{\normalfont\footnotesize \noindent\unhbox\@tempboxa#2}%
	\else
	\hbox to\hsize{\normalfont\footnotesize\hfil\box\@tempboxa\hfil}\fi\fi}
\makeatother

\usepackage{balance}
\usepackage{cite}
\usepackage{times}
\usepackage{graphicx}
\usepackage{color}
\usepackage{ifpdf}
\usepackage{subfigure}
\usepackage{epsfig}
\usepackage{latexsym}
\usepackage{amsfonts}
\usepackage{amssymb}
\usepackage{comment}
\usepackage{xspace}
\usepackage{mathrsfs}
\usepackage{amssymb}
\usepackage{color}
\usepackage{setspace}%
\usepackage{algorithmic}
\usepackage[ruled]{algorithm}
\usepackage{amssymb}
\usepackage{url,epsfig,array}
\usepackage{leftidx}
\usepackage{amsmath}

\def\ie{\textit{i.e.}\xspace}

\newtheorem{theorem}{Theorem}

\newtheorem{definition}{Definition}

\newcommand{\argmax}{\operatornamewithlimits{argmax}}

\ifodd 0
\else

\fi

\begin{document}

\begin{spacing}{1.5} 
%
\title{TARCO: Two-Stage Auction for D2D Relay Aided Computation Resource Allocation in Hetnet\thanks{Working paper, submitted to IEEE Transactions on Services Computing, neighter the revised version nor final version }}

{
	\author{\IEEEauthorblockN{Long Chen\IEEEauthorrefmark{1}, Jigang Wu\IEEEauthorrefmark{2} 
		, Xinxiang Zhang}  \\
		\IEEEauthorblockA{
			School of Computer Science and Technology, Guangdong University of Technology,  China\\
			\IEEEauthorrefmark{1}lonchen@mail.ustc.edu.cn,\IEEEauthorrefmark{2}asjgwucn@outlook.com 
			}}
}

\markboth{SUBMITTED to IEEE TRANSACTIONS ON SERVICES COMPUTING}{}
\maketitle

\section{abstract}
In heterogeneous cellular network, task scheduling for computation offloading is one of the biggest challenges. Most works focus on alleviating heavy burden of macro base stations by moving the computation tasks on macro-cell user equipment (MUE) to remote cloud or small-cell base stations. But the selfishness of network users is seldom considered. Motivated by the cloud edge computing, this paper provides incentive for task transfer from macro cell users to small cell base stations. The proposed incentive scheme utilizes small cell user equipment to provide relay service. The problem of computation offloading is modelled as a two-stage auction, in which the remote MUEs with common social character can form a group and then buy the computation resource of small-cell base stations with the relay of small cell user equipment. A two-stage auction scheme named TARCO is contributed to maximize utilities for both sellers and buyers in the network. The truthful, individual rationality and budget balance of the TARCO are also proved in this paper. In addition, two algorithms are proposed to further refine TARCO on the social welfare of the network. Extensive simulation results demonstrate that, TARCO is better than random algorithm by about $104.90\%$ in terms of average utility of MUEs, while the performance of TARCO is further improved up to $28.75\%$ and $17.06\%$ by the proposed two algorithms, respectively.

\begin{IEEEkeywords}
	\emph{Auction, Computation Offloading, HetNet, incentive, Relay.}
\end{IEEEkeywords}

\section{Introduction}\label{sec:intro}
With the proliferation of wireless communication devices such as smartphones, tablets, laptop computers and so on, both spectrum and data transmission rate demands are becoming much higher. According to the latest Cisco visual networking index report, the global mobile data traffic will be about 30.6 exabytes per month by 2020 \cite{indexcisco}. To accommodate such urging demands, deploying small cells underlaying macro cells as well as device-to-device (D2D) communications are the dominant strategies for the emerging next generation wireless communications (5G). By deploying low transmission power small cells overlaid macro cells, network coverage and capacity can be improved \cite{zhu2014pricing}. Meanwhile, D2D communication in a small cell can further enhance the spectrum utilization efficiency via spectrum reuse such that the interferences between D2D transmission pairs as well as between D2D pair and small cell user are well controlled. Therefore, D2D based heterogeneous networks (HetNets) have been a hot research topic in recent years.

The prior art in D2D based HetNet mainly focused on coverage expansion \cite{jutruthful}, power control \cite{xing2010investigation}, interference management \cite{yin2015pricing}, resource allocation \cite{song2014game,malandrino2015interference} and so on. As an effective solution to handle channel fading and increase transmission rate in wireless communications, relay aided communication has been widely used, including two-hop or multi-hop relaying and two-phase cooperative transmission \cite{shi2008optimal}. Therefore, relay aided communication in D2D based HetNet can also improve the network performance. For example, by hiring a small cell user equipment (SUE), the macro cell base station (MCB) can reach to the macro cell user equipment (MUE) in the downlink transmission, thus the coverage of the macro cell can be extended \cite{ur2015analysis}. Authors in \cite{xiao2015optimal,xiao2016energy} optimized the system spectrum efficiency and energy efficiency via D2D relay as well as modes selection. A load balancing scheme for D2D-based relay communications in HetNet is proposed by \cite{chen2015load}. A small cell user relay aided downlink communication from macro base station to remote macro cell user is proposed via D2D communication \cite{hwang2015ue}. In the D2D based HetNet architecture, there are a few papers that address incentive mechanisms. Authors in \cite{jiang2015rally} proposed a D2D based content sharing game to maximize cellular offloading. Zhang et al. \cite{zhang2015contract} came out with a contract theory based scheme to attract user equipment delivering data for the cellular base station.

Recently, Mobile-Edge Computing (MEC) has been proposed and it is an emerging paradigm to improve the quality of experience (QoS) of mobile devices at the edge of pervasive radio access networks \cite{chen2015efficient}. In \cite{sardellitti2015joint}, a computation offloading mechanism for mobile users was implemented by offloading computation tasks to a common server in HetNet. It is noteworthy that, in real applications, mobile devices are carried by human with limited budget. A single resource user may not afford to buy the computation resources on a single SCB. For example, a mobile user may just want to offload the translation task consisting of a single phrase, however, at that time the SCB may want to lease out a whole computation resources larger than finishing the simple translation task. In this case, the charging price of SCB may be very high and it may not be motivated to execute the task due to small monetary revenue. In real scenarios, although different mobile user may have different application level computation tasks, people in the same area may have the same level social relations \cite{semiari2015context} or perform similar tasks, they can form a small scale social group \cite{cao2016share} to buy the computation services offered by the SCB. Group based auction \cite{Lin2013Groupon} has been proposed and is proved to be an effective mechanism to collect small bids and hiring a bid agent to buy resources from resource owners. When more users are willing to participate in the group based resource sharing, the benefits of resource owners will also increase and the resource utilization efficiency will be enhanced. Meanwhile, the revenue of individual buyers can be enlarged due to price discount \cite{chen2015spectrum}. 

On the computation offloading problem in D2D based HetNet, it is noteworthy that, part of works \cite{ur2015analysis,xiao2015optimal,xiao2016energy,chen2015load,hwang2015ue} are under the asumption that users are willing to cooperate, which is far from realistic. In this paper, we employ auction theory to provide incentive for mobile users. Some works \cite{jiang2015rally,zhang2015contract,chen2015efficient,sardellitti2015joint} do not take relay influence into consideration. It should be noted that when mobile users are far from the target server, direct communication may not ensure the communication consistency and user QoS may suffer. Motivated by the recent work \cite{jo2015device} for mobile computing where master device is chosen to offload the computation tasks of mobile users via D2D communication in HetNet, in this paper, we provide incentive for idle SUEs within small cells to act as relays by offloading computation tasks of MUEs to small cell base stations (SCBs), i.e. the edge servers. Moreover, most of existing researches on HetNet focus on maximizing the revenue of the network infrastructure to optimize the load of base stations, instead of mobile users. Note that computation tasks and resource demands are generally heterogeneous in real application. Thus, it is difficult to guarantee the efficiency of mobile users, only by optimizing the base station's traffic burden. Therefore, it is significant to investigate new network paradigm on the existing two-tier HetNet archietecture. Inspired by the above mentioned group based auction, this paper tries to group the MUEs in low signal area or far from the MCB, and helps to select proper idle SUEs for each MUE group. Different from \cite{Lin2013Groupon} and \cite{yang2014truthful}, in this paper, we design the scheme tailored for relay aided coputation offloading taking account of user's demands, where user's demands are not properly addressed in \cite{Lin2013Groupon} and \cite{yang2014truthful}. 

To effectively offload computation tasks of MUEs and design new mechanisms, we have to solve the following challenges: (i) Running applications on MUEs are heterogeneous, such as text translation, video decoding, machine learning tasks etc. Therefore, the budget and demand for each MUE are different. Meanwhile, based on the distances from MUEs to SCBs, the valuation of each MUE for the computation capability of SCBs is various. How to determine winning prices to attract MUEs is a challenge. (ii) From the SCB's perspective, offloading MUEs' tasks will consume its power and computation cycles. Meanwhile, for idle SUEs, their service for relaying MUEs should also be motivated because of the selfishness nature. Therefore, how to provide monetary compensation should be carefully designed for both SUEs and SCBs. (iii) Auction should be truthful, budget balance, individual rational and efficient. How to ensure the above characteristics at the same time in a two-stage auction is a challenge. 

The main contribution of this paper can be summarized as follows:
\begin{itemize}
	\item Different from traditional auctions, in this paper, the social relations among different MUEs and the similarity between tasks are considered where MUEs can form a group with similar tasks or level of social ties to buy computation resources from SCBs.
	\item To handle the selfishness nature of entities in the network and provide incentive for them, a group based \textbf{t}wo-stage \textbf{a}uction for \textbf{r}elay aided \textbf{c}omputation \textbf{o}ffloading in HetNet (\textbf{TARCO}) is proposed.
	\item We prove the proposed scheme guarantees three important economic properties: truthful, budget balance and individual rationality. 
	\item To enhance the performance of TARCO, we design MWD and VITA algorithms to improve the social welfare of the network. Extensive simulation results demonstrate that TARCO is about $104.90\%$ better than RND algorithm on average utility of MUEs and VITA is $28.75\%$ better than TARCO and MWD outperforms TARCO for about $17.06\%$. 
\end{itemize}

The remainder of this paper is organized as follows. Section \ref{sec:relate} discusses related works on task offloading in the HetNet, both non-incentive and incentive based works have been compared. Section \ref{sec:sysmodel} presents the system model. Section \ref{sec:auction} shows the detail of two-stage auction scheme. In Section \ref{sec:analysis} we prove the economic properties of the auction scheme. Then we propose improved schems to enhance the social warefare of the system in Section \ref{sec:improve}. Section \ref{sec:simulation} gives performance evaluation results and finally Section \ref{sec:conclude} concludes this paper.

%
\section{Related Work}\label{sec:relate}
Recently, there has been related work on user association, computation or traffic offloading in heterogeneous wireless networks. Both non-incentive based and incentive based schemes have been proposed. In the following, we summarize and compare the related work on the above categories. 

\subsection{Non-incentive based schemes}
Some works study D2D based user association in HetNet \cite{xiao2016energy} \cite{liu2015device}. Xiao et al. \cite{xiao2016energy} proposed an energy efficient mode selection and user association scheme. The MCB acts as a central controller to determine which one of SCB and MCB that the mobile device should be attached to. Meanwhile, D2D relay mode and direct communication mode selection is also considered. Liu et al. \cite{liu2015device} adopted a D2D two-hop relay to help mobile users get access to neighbouring base stations. In this paper, tasks on MUEs in the two-tier HetNet are offloaded to SCBs instead of the far away MCB to conquer the high channel fading and low signal obstacles of MUEs. 

Rahman et al. \cite{ur2015analysis} considered the downlink coverage extension from the MCB to MUEs via the D2D cooperation between SUEs and MUEs in a single macro cell. In this paper, we focus on the uplink transmission by offloading the computation task from MUEs to SCBs. The most recent work by Cao et al. \cite{cao2016cellular} designed a hybrid traffic offloading scheme for HetNets where mobile users can get access to SCBs with the help of mobile relays via D2D communication. Their objective is to maximize the number of mobile users admitted into the whole network, through MCB or SCBs. The problem is formulated as an integer linear programming and solved by dynamic programming. Different from \cite{cao2016cellular}, this paper mainly focuses on MUE's tasks offloading to the SCB with help of SUE relays. The benefits of both SUEs and SCBs can be ensured by the Tarco scheme. In \cite{kawamoto2014efficient}, a heuristic method was proposed to detour congested MCB traffic to uncongested SCBs with the help of D2D relay nodes. However, only skeletal numerical analysis is given by \cite{kawamoto2014efficient}, which cannot be directly deployed. Thus we carry out with a distributed implementation.

\subsection{Incentive based mechanisms}
Recently, the incentive based schemes have been proposed. Hua et al. \cite{hua2013truthful} proposed a truthful auction framework for femtocell access between MUEs and femtocell base stations. A dynamic game is designed by Zhu et al. \cite{zhu2014pricing} to offload traffic from the MCB to SCB service providers. However, both in \cite{hua2013truthful} and \cite{zhu2014pricing}, relay is not used. 

LeAnh et al. \cite{leanh2015joint} came out with a Stackelberg game based offloading mechanism by transferring part of MUE's data to the SCB via SUE relay nodes. The MUE is the leader while candidate SUE relays are followers. The optimal power allocation and relay selection is conducted via leader-follower pricing. Later, Ho et al. \cite{ho2016coordinated} proposed a two-stage Stackelberg game for SCBs to admit MCB's traffic from the perspective of network fixed infrastructure, however, relay influence is not considered. Meanwhile, they do not consider SCB's heterogeneity.

Methods proposed by other literatures cannot be applied in this paper. Yang et al. \cite{yang2014truthful} studied the spectrum group auction in cognitive radio networks. Later, Wang et al. \cite{wang2015truthful} came out with a scheme to allocate both channels and cooperative relay nodes for cooperative cognitive radio networks with two-stage group auction. However, in their work, the ask price at each spectrum owner is same for different buyers. In computation offloading, the tasks are different and for each SCB, the ask price may not be constant for all buyers due to different location and channel fading conditions. In mobile cloud computing, Jin et al. \cite{jin2016auction} carefully designed an incentive-compatible auction scheme between cloudlets and mobile users. The cloudlets are edge servers to offload traffic of remote cloud. Similar to \cite{jin2016auction}, the SCBs can be regarded as cloudlets to admit computation tasks sent from mobile users.  

Based on the observations, to offload the computation tasks of MUEs with relay SUEs and motivate both SUEs and SCBs should be properly handled. New schemes should deal with heterogeneous computation offloading tasks as well as SCB's heterogeneity. 
\section{System Model and Problem Formulation}\label{sec:sysmodel}
In this section, firstly we present the system model, then the computation offloading problem of MUEs is formulated as a two-stage auction. At last, we introduce related economic properties that auction scheme should follow. The basic notations are shown in table \ref{tb:nt}.

\subsection{Main Idea of TARCO Scheme}
\begin{figure}[ht]
  	\centering
   	\includegraphics[width=3.5in]{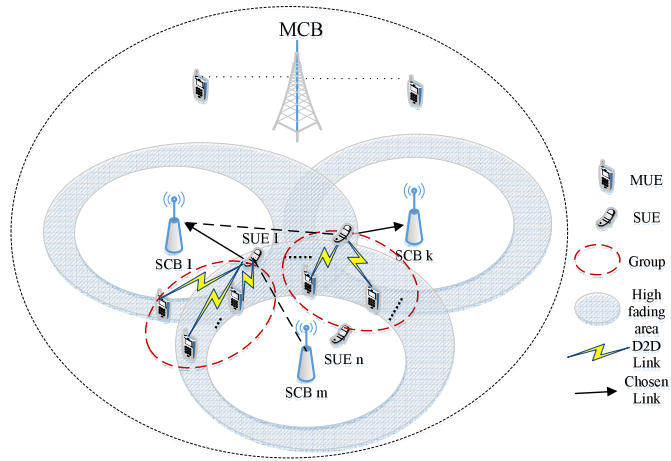}
   	\caption{Network Model}\label{fig:network}
\end{figure}

The main idea of our proposed TARCO scheme is to provide incentive and utilize D2D communications to assist the computation offloading of MUEs to local SCBs, with the aim to maximize utilities of both SCBs and MUEs. Take Fig. \ref{fig:network} as an example, when there are some MUEs in the high channel fading area or low signal region, they cannot directly to offload their computation tasks to neighbour SCBs without the help of SUE relays. The proposed TARCO scheme allows MUEs to access the SCBs by relaying the computation data via SUE relays in the coverage of corresponding SCBs. Intuitively, TARCO can improve the system performance by reducing the traffic flowed to MCB and reduce the MUE's energy consumption with a lower transmission power. It should be noted that for MUEs with high mobility and moving constantly, it may get access to MCB to avoid frequent handover, which is out of the scope of this paper.

\subsection{System Model}\label{subsec:sysmodel}

\begin{table}[t]
	\centering
	\small
	\caption{Notations for system model}\label{tb:nt}
	\begin{tabular}{l|p{6.5cm}}
		\hline
		Notations       & Meaning                                                    \\ \hline
		$s_i^{j}$, $S_i$ & The $j$th MUE in the $i$th SUE group, the $i$th group       \\
		$N$, $n_i$      & Number of SUEs and MUEs in the $i$th  SUE group               \\
		$b_i^j(k)$, $v_i^j(k)$  & True bid and valuation of $s_i^{j}$ for the $k$th SCB  \\
	    $d_i^j(k)$     & Demand of $s_i^{j}$ for the $k$th SCB                            \\	
	    $\bar{b}_i^j(k)$, $\bar{v}_i^j(k)$	 &  Bid and valuation of $s_i^{j}$ for the $k$th SCB \\
	    $p_i^j(k)$, $P_i$ & $s_i^j(k)$'s and the $i$th SUE's payment for the $k$th SCB  \\
	    $B_i^k$ & The $i$th SUE's bid for the $k$th SCB  \\
	    $F_i(k)$ & The total payment of the $i$th SUE for the $k$th SCB　　\\
	    $A_k$       & Ask or reserve price of the $k$th SCB    \\
	    $p_c$     & The clearing price  \\
	    $u_i^j$   & The utility of MUE $s_i^j$ \\
	    $U_i$     & The utility of the $i$th SUE $r_i$ \\
	    $U_k$     & The utility of the $k$th SCB $e_k$  \\
	    $\mathcal{E}_w$, $\mathcal{R}_w$  & Winning set of SCBs, Winning set of SUEs \\
	    $S_i^w$  &  Winning set of MUEs for SUE $r_i$  \\
	    \hline
	\end{tabular}
\end{table}

In our scenario, we assume all nodes stay static in a given auction period. In the network shown in Fig. \ref{fig:network}, there is one MCB and $M$ SCBs in the coverage of the MCB and neighbouring SCBs have orthogonal frequency bands \cite{liu2016energy}. Meanwhile, MUEs are heterogeneous and need to perform different computation tasks. Each offloading task contains a QoS requirement information such as bandwidth and CPU cycles. Each MUE may have its own preference for the SCB based on its geological position, channel condition and distance from the SCB. Since MUEs within the same location may have similar relationships \cite{semiari2015context} and sense of channel conditions, they may get-together and form a communication social group. In total, there are $N$ groups of MUEs. By adopting interference mitigation technologies \cite{khawer2016usicic}, interferences between adjacent groups can be properly handled. Thus in this model, we do not consider the interference between small cells.  Based on computation tasks, for the benefit of simplicity, we assume each MUE may require different number of CPU cycles. In the $i$th social group $S_i$, where $i\in [1,N]$, there are $n_i$ MUEs and $S_i=\{s_{i}^{1},s_{i}^{2},\cdots,s_{i}^{n_i}\}$. We use $\mathcal{E}=\{e_1,e_2,\cdots,e_M\}$ to denote the set of SCBs, i.e. the edge servers and $\mathcal{R}=\{r_1,r_2,\cdots,r_N\}$ to denote the set of SUEs. The $k$th SCB is denoted by $e_k$, $k\in [1,M]$. For each SCB $e_k$, each MUE $s_{i}^{j}$ has a CPU cycle demand $d_{i}^{j}(k)$, its valuation $v_{i}^{j}(k)$ and the budget $b_{i}^{j}(k)$ for the maximum payment. For the benefit to express the preference of MUE $s_i^{j}$ towards $e_k$, we define
\begin{equation}\label{eq:vijkdef}
v_{i}^{j}(k)=\max\{\frac{C_R(s_{i}^{j},r_{i},e_k)}{C_D(s_{i}^{j},e_k)},0\}. 
\end{equation}

It should be noted that the CPU cycle demand may be different for different SCBs because of the location of SCBs and the traffic burden on them.

We design a two-stage hierarchical auction for the computation offloading. TARCO consists of two single round sealed bid auctions, they are tier I auction and tier II auction accordingly. The tier I auction is between each social group $S_i$ and SUE relay $r_i$. Each SUE relay $r_i$ is in charge of all the members in $S_i$. The tier II auction is between SUE $r_i$ and SCB $e_k$. Let $B_{i}^{k}$ be the $r_i$'s bid for SCB $e_k$, $i\in [1,N], k\in [1,M]$. Then the ask price of $e_k$ is $A_k$, which is the reserve price. We assume each social group can be served by one SCB at most because serving for multiple social groups will cause longer delays. Meanwhile, we also assume, each members inside the group will choose only one SCB.

In tier I aucton, for SCB $e_k$, each member of $S_i$ submits the bid $\beta_{i}^{j}(k)=\{{b}_{i}^{j}(k),{v}_{i}^{j}(k), d_{i}^{j}(k)\}$ to $r_i$. Note that $s_i^{j}$ may not be honest and report its true bid, therefore we have $\bar{\beta}_{i}^{j}(k)=\{\bar{b}_{i}^{j}(k),\bar{v}_{i}^{j}(k), d_{i}^{j}(k)\}$ if it may bring higher utility. For the SUE $r_i$, after gathering the group member's bids, the SUE $r_i$ will determine winning MUEs in $S_i(k)$, which is denoted by $S_i^{w}(k)$ and $S_i^{w}\subseteq S_i$. If $s_i^{j} \in S_i^{w}(k)$, and $e_k$ is selected by $r_i$, then $s_i^{j}$ will be charged $p_i^{j}(k)$. For $r_i$, the gathered bid is $F_i(k)$. In tier II, $r_i$ submits its bid for SCB $e_k$ as $B_i^{k}$, where $i\in [1,N], k\in [1,M]$. A double auction will be conducted to determine how many CPU cycles to allocate to relay SUEs and the payments for winners. If the SUE fails to obtain resources from the SCB, no transactions will happen and the utility of MUE members in the group will be zero.

Given the above settings, we denote $\mathcal{E}_w \subseteq \mathcal{E}$ as winning set of SCBs and $\mathcal{R}_w \subseteq \mathcal{R}$ as the winning relay set in tier II auction. Let $P_i$ be the price the wining SUE $r_i$ needs to pay for the SCB. If SUE $r_i$ wins then the payment of $r_i$ should not be greater than its gathered bid $F_i(e_i^{w})$, where $e_{i}^{w}\in \{0,1,2,\cdots,M\}$.

\subsection{Problem Formulation}
Let $u_{i}^{j}$ denote the utility of MUE $s_{i}^{j}$, for each MUE $s_{i}^{j}\in S_{i}^{w}$, SUE $r_i$ computes the payment $p_{i}^{j}$. Note that the payment for the winning MUE $s_i^{j}$ should not be higher than the true budget $b_{i}^{j}$. Hence the utility of MUE $s_i^{j}$ is

\begin{equation}\label{eq:uij}
u_i^{j}=
\begin{cases}
 v_i^{j}(k)-p_i^{j}(k),   &\text{if $s_i^{j} \in S_i^{w}$ and $p_i^{j}(k)\le b_{i}^{j}(k)$} \\
 0,                 &\text{otherwise}.\\
\end{cases}
\end{equation}
Where we have $p_i^{j}(k)$, which is defined as \\
\begin{equation}\label{eq:pijk}
p_i^{j}(k)=p_c^{i}(k) d_i^{j}(k), \text{$k=e_i^{w}$ if and only if $s_i^{j} \in S_i^{w}$},\\
\end{equation}
where $p_c^{i}(k)$ is the clearing price. Accordingly, the utility of SUE $r_i$ is defined as\\
\begin{equation}\label{eq:Ui}
U_{i}=
\begin{cases}
F_i(k)-P_i, &\text{if $r_i \in \mathcal{R}_w$, $e_k \in \mathcal{E}_w$}\\
0, &\text{otherwise}.
\end{cases}
\end{equation}

Further more, the utility $U_k$ of SCB $e_k$ should not be negative, which is \\
\begin{equation}\label{eq:Uk}
	U^k=
\begin{cases}
P_i-A_k, &\text{if $e_k \in \mathcal{E}_w$ and $r_i$ wins $e_k$} \\
0, & \text{otherwise}.
\end{cases}
\end{equation}

The first branch of (\ref{eq:Uk}) is the utility of $e_k$ when it wins the auction and gets enough payment from auctioneer in the transaction. The utility of $e_k$ should not be negative when it offloads computation tasks from MUEs.

\subsection{Economic Properties}\label{sec:subsececonoprop}
An auction will not be conducted until many economic properties are satisfied. In the following paragraphs, several critical economic properties of the auction are listed which TARCO would like to achieve.

\begin{definition}\label{def:truth}
(Truthful). An auction is truthful if any participant's utility is maximized by submitting its true valuation, regardless what others would like to behave. That is, no bidder can improve his utility by misreporting his bids.
\end{definition}

\begin{definition}\label{def:indivrat}
	(Individual Rationality). The utility of each participant of the auction is non-negative. That is, $u_i^{j}$, $U_i$ and $U_k$ are non-negative in the TARCO auction scheme.
\end{definition}

\begin{definition}\label{def:budget}
	(Budget Balance). An auction is budget balanced if total payment from buyers are no less than the sum paying paid to sellers. In TARCO scheme, this property is required in tier II auction to ensure the auctioneer a non-negative utility, \ie $\sum_{r_i\in R_w} P_i \le \sum_{e_k\in \mathcal{E}_w} P_k$.
\end{definition}

\begin{definition}\label{def:efficiency}
(Computational Efficiency). An auction is computational efficiency if the scheme can terminate in polynomial time in terms of the input.
\end{definition}

In this paper, TARCO is designed to achieve truthful, individual rationality, budget balance and computational efficiency. The mechanism is illustrated in detail in Section \ref{sec:auction} followed by proves in Section \ref{sec:analysis}. 
\section{Two-Stage Auction Mechanism}\label{sec:auction}
In this section, we propose a two stage based auction scheme for relay aided computation offloading scheme in HetNet, which is called TARCO. TARCO satisfies properties such as truthful, individual rationality, budget balance and computational efficiency proposed in Section \ref{sec:subsececonoprop}.

Specifically, TARCO consists of two sub-auctions, the tier I auction and the tier II auction. In tier I auction, all MUEs who need to offload their computation tasks, submit their bids consisting of valuation, budget and demands to the corresponding SUE they attach to. The tier I auction is conducted virtually by each SUE node $r_i\in \mathcal{R}$ and will not be executed until $r_i$ wins the following procedures. In tier II auction, each SUE submits its bid to the MCB based on the collected payment paid by MUEs in tier I auction. If the SUE $r_i$ wins a SCB, it will charge the winners in its group $S_i$ and relay the computation traffic to the SCB.

\subsection{Procedure 1: Tier 1 Auction}
In the procedure, the relay SUE $r_i$ will conduct the auction virtually and decide winners. First, each SUE obtains the characteristics of computation abilities from SCBs, such as the number of available CPU cycles, remained energy, and geological information. These properties will be sent to MUEs with the help of SUE relays aiming to assist the decision made by MUEs. The SUE relay will receive the three-dimension bids for computation services for its MUEs which is demoted by $\{b_i^{j}(k),v_i^{j}(k),d_i^{j}(k)\}$. Let $OPT(b/d)$ be the unit budget of the optimal single-price auction denoted by
\begin{equation}\label{eq:bd}
OPT(b/d)=\max_{1\le i \le |b|} i\frac{b_i}{d_i},
\end{equation}\\
where $|b|$ denotes the length of the array, $b_i$ denotes the $i$th budget and $d_i$ denotes the $i$th demand.

In this stage, the designed algorithm should calculate the budget vector $F_i(k)$ for SCB $k$. We compute the budget for each SUE based on the random sampling profit extraction auction \cite{fiat2002competitive} and inspired by the work in \cite{goldberg2001competitive}, we partition the MUE set $S_i$ into two sets, $S_i^{1}$ and $S_i^{2}$ uniformly for sampling purpose. Then SUE $i$ computes $R^{1}=OPT(b^1/d^1)$ and $R^{2}=OPT(b^2/d^2)$. Depending on the values of $R^1$ and $R^2$, the budget of MUE is extracted from both vectors. The detail of the algorithm is shown in Algorithm \ref{alg:comptbgt} and \ref{alg:ph1}.

\begin{algorithm}[h]
	\begin{algorithmic}[1]\caption{ComptBgt $(b/d,R,(v_i^{j}(k))_{j=1}^{n_i})$}\label{alg:comptbgt}
		\REQUIRE{Sorted vector of b/d, potential budget R and valuation vector $(v_i^{j}(k))_{j=1}^{n_i}$}
		\ENSURE{Budget with given valuation}
		\STATE{Search for the maximum $j$ in $b/d$ such that $jb_j/d_j \ge R$} 
	    \STATE{$p_c\leftarrow \frac{R}{j}$}
	     \STATE{$S_i^{w}(k)\leftarrow \emptyset$}
	     \STATE{$F_i(k)\leftarrow \emptyset$}
	     \FOR{$j\leftarrow 1$ to $n_i$}
	     \STATE{$p_i^{j}(k)\leftarrow p_c\cdot d_i^{j}(k)$}
	     \IF{$p_i^{j}(k)<b_i^{j}(k)$ and $p_i^{j}(k)<v_i^{j}(k)$}
	     \STATE{$S_i^w(k)\leftarrow S_i^{w}(k) \cup s_i^{j}$}
	     \STATE{$F_i(k)\leftarrow F_i(k)+p_i^{j}(k)$}
	     \ENDIF
	     \ENDFOR
	     \STATE{return $F_i(k)$}
		\end{algorithmic}
	\end{algorithm}

\begin{algorithm}[h]
	\begin{algorithmic}[1]\caption{Phase I of TARCO for SUE $r_i$}\label{alg:ph1}
		\FOR{$k\leftarrow 1$ to $K$}
		\STATE{Let $b/d$ denote the sorted aray of $(b_i^{j}(k)/d_i^{j}(k))$ in non-increasing order}
			\STATE{Divide $b/d$ uniformly at random into two arrays $b^1/d^1$ and $b^2/d^2$}
			\STATE{$R^1 \leftarrow OPT(b^1/d^1), R^2 \leftarrow OPT(b^2/d^2)$}
			\IF{$R^1<R^2$}
			\STATE{$F_i(k)\leftarrow$ ComptBgt($b^2/d^2,R^1,(v_i^{j}(k))_{j=1}^{n_i}$)}
			\ELSE
			\IF{$R^1>R^2$}
			\STATE{$F_i(k)\leftarrow$ ComptBgt($b^1/d^1,R^2,(v_i^{j}(k))_{j=1}^{n_i}$)}
			\ELSE
			\STATE{$F_i(k)\leftarrow$ [ComptBgt($b^2/d^2,R^1,(v_i^{j}(k))_{j=1}^{n_i}$)$+$\\ComptBgt($b^1/d^1,R^2,(v_i^{j}(k))_{j=1}^{n_i}$)]/2}
			\ENDIF
			\ENDIF
			\ENDFOR
		\end{algorithmic}
	\end{algorithm}

From the view of relay $r_i$, $s_i^{j}$ can be regarded as the buyer with $d_i^{j}(k)$ computation resource to buy on budget $b_i^{j}(k)$. According to Algorithm \ref{alg:comptbgt}, the clearing price of $s_i^{j}$ for the $k$th SCB should not be greater than the unit budget $b_i^{j}(k)/d_i^{j}(k)$. Note that the value $R$ is extracted from smaller value group in the partition procedure of Algorithm \ref{alg:ph1} to ensure truthfulness. The decision process for winner set $S_i^{j}(k)$ and the gathering of $F_i(k)$ is regarded as the start for $r_i$ to bid SCB $k$ in the second phase, \ie phase II. The winners in $S_i(k)$ will get the desired CPU cycles if and only if SUE $r_i$ wins SCB $e_k$ in phase II.

\begin{figure}[h]
\includegraphics[width=3.5in]{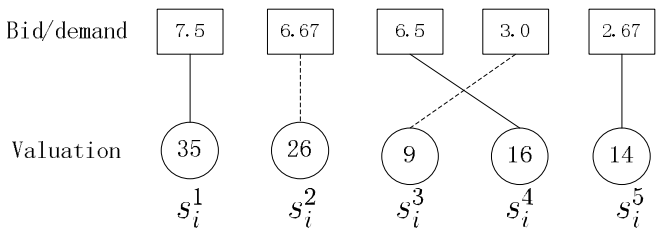}
\caption{Example of Algorithm \ref{alg:ph1}, where the dashed lines belong to $b^2/d^2$ and the remaining lines belong to $b^1/d^1$}\label{fig:example}
\end{figure}

To illustrate how Algorithm \ref{alg:ph1} works, we propose a simple example as follows. There are $5$ MUEs with their bids for SCB $e_k$ as $(30,4,35)$, $(20,3,26)$, $(18,6,9)$, $(13,2,16)$, $(8,3,14)$, where in each pair the first number is the bid, the second item is number of demand and the third item is the valuation, hence $b/d=\{7.5, 6.67, 3, 6.5, 2.67\}$. Assume that $b^1/d^1=\{7.5,6.5,2.67\}$, $b^2/d^2=\{6.67,3\}$ and we have $R^{1}=13$, $R^{2}=6.67$. Since $R^{2}<R^{1}$, we then extract $R^{2}$ from $b^2/d^2$. The mapping between the sorted value $b/d$ and valuation is illustrated in Fig. \ref{fig:example}. Obviously, it is shown that in Fig. \ref{fig:example}, for MUEs in group $i$, the clearing price $p_c^{i}(k)$ for SCB $k$, which is $p_c^{i}(k)=6.67/3=2.223$. Since the value $p_i^{1}(k)=p_c^{i}(k)\times d_{i}^{1}(k)=2.223\times 4=8.892<b_i^{1}(k)=30$ and $p_i^{1}(k)<v_i^{1}(k)=35$. Therefore, $s_i^{1}$ is added to the winner set $S_i^{w}(k)$ for the $k$th SCB. Similarly, $s_i^{2}$, $s_i^{4}$, $s_i^{5}$ are all winners. Therefore, the total budget of $r_i$ obtained from group $S_i$ for SCB $e_k$ is $26.676$.

\subsection{Procedure 2: Tier 2 Auction}
After the previous procedure, \ie phase I of auction which is conducted between MUEs and relay SUEs, the budget for SUE $r_i$ bids for SCB $e_k$ can be obtained and denoted as $F_i(k)$. $F_i(k)$ will be gathered when the SUE $r_i$ wins SCB $e_k$. In procedure 2 of the auction, there are multiple sellers and multiple buyer with heterogeneous items to bid since each SUE has different budgets for the corresponding SCBs. In this research, VCG auction is not adopted because of the high computational cost and the failure to ensure budget balance property. Meanwhile, the McAfee auction cannot be utilized since it only suits for the scenario where there are homogeneous goods to trade \cite{yang2011truthful}. Therefore, new auction mechanism should be invented in this procedure. Here, we propose a Random Matching based Efficient Allocation algorithm RMEA to trade the computation resource between SUEs and SCBs.

The detail of RMEA is shown in Algorithm \ref{algo:ph2}. The SCB assigns CPU cycles to each SUE relay node in a greedy manner. For each SUE relay $r_i$, the auctioneer try to maximize $B_i^{k}-A_k$. Note that the auctioneer always choose the minimum index of $k$ when there are multiple choices. The procedure resembles fixed price auction, which is proved to be truthful \cite{goldberg2001competitive}. RMEA charges each winning SUE $r_i$ the amount of $P_i$ and pays each winning SCB $e_k$ the payment $P_i$. 
\begin{algorithm}[ht]
	\begin{algorithmic}[1]\caption{Phase II of TARCO, the RMEA algorithm}\label{algo:ph2}
		\REQUIRE{$B_i^{k}$, for $\forall 1\le i \le N$ and $1\le k \le M$}
		\ENSURE{${\mathcal{R}}_w$, ${\mathcal{E}}_w$, $P_i$}
		\STATE{$\mathcal{E}\leftarrow \{e_1,e_2,\cdots, e_M\}$, $\mathcal{E}_w \leftarrow \emptyset$, $\mathcal{R}_w \leftarrow \emptyset$}
		\FOR{$i\leftarrow 1$ to $N$}
		\STATE{$e_k \leftarrow \argmax_{e_k \in \mathcal{E}}(B_i^{k}-A_k)$}
		\IF{$B_i^{k}-A_{k}\ge 0$}
		\STATE{$\mathcal{R}_w \leftarrow \mathcal{R}_w \cup \{r_i\}$}
		\STATE{$\mathcal{E}_w \leftarrow \mathcal{E}_w \cup \{e_k\}$}
		\STATE{$P_i \leftarrow B_i^{k}$}
		\STATE{$\mathcal{E}\leftarrow\mathcal{E}\setminus \{e_k\}$}
		\ENDIF
		\ENDFOR
	\end{algorithmic}
\end{algorithm}

\subsection{Charge for MUEs}\label{subsec:charge}
As mentioned in previous subsection, the phase I of TARCO is conducted virtually by each SUE node and the auction results will not be executed unless the SUE wins the phase II auction. Next, we give a simple example to illustrate the charging procedure.

Assume there are $3$ SCBs and $3$ SUEs. At the end of phase I, we may have $F_1(k)=3$, $F_2(k)=5$, $F_3(k)=7$, for $\forall k\in [1,3]$. In phase II,  $B_1^{k}=2$, $B_2^{k}=5$, $B_3^{k}=4$ and accordingly the ask price of SCBs are $A_1=1$, $A_2=3$, $A_3=5$. After the execution of algorithm 
\ref{algo:ph2}, the winning SUE relays are $r_1$ and $r_2$, accrodingly, the winning SCBs are $e_1$ and $e_2$.  Therefore, the utilities of SUEs are $U_1=F_1(k)-B_1^{k}=3-2=1$, $U_2=F_2(k)-B_2^{k}=5-5=0$ and the utilities of SCBs are $U^{1}=P_1-A_1=2-1=1$, $U^{2}=P_2-A_2=5-3=2$. Note that, the bid value of $B_i^k$ may be lower than $F_i(k)$. 

\section{Theoretical Analysis}\label{sec:analysis}
In this section, we focus on proving the economic properties of proposed TARCO scheme. It is shown that TARCO satisfies all the following properties mentioned in previous section, which is given in theorems below.

\begin{theorem}\label{theorem:truthMUE}
TARCO is truthful for MUEs in the network.
\begin{proof}
We will prove that the dominant strategy for the MUE $s_i^{j}(k)$, whose bid vector is $\{b_i^{j}(k),v_i^{j}(k),d_i^{j}(k)\}$, is to submit its bid as its true bid, which is $b_i^{j}(k)=\bar{b}_i^{j}(k)$, $v_i^{j}(k)=\bar{v}_i^{j}(k)$ and $d_i^{j}(k)=\bar{d}_i^{j}(k)$. It should be noted that the demand value $d_i^{j}(k)$ is truthful because it reflects the true demand of the MUE $s_i^{j}(k)$. 

Assume that $s_i^{j}$ is in the subset $S_i^{1}$, then the proof for $s_i^{j}$ in subset $S_i^{2}$ is similar. From the first branch of equation (\ref{eq:uij}), we can see that $s_i^{j}$ can improve its utility by manipulating its reported valuation $v_i^{j}(k)$ and budget $b_i^{j}(k)$. However, on the hand, we can see that the clearing price $p_c^{i}(k)$ is determined independently of $s_i^{j}$'s budget $b_i^{j}(k)$ and valuation $v_i^{j}(k)$ in Algorithm \ref{alg:comptbgt} and Algorithm \ref{alg:ph1}. Since the clearing price is based on random sampling auction \cite{fiat2002competitive}, the truthfulness for procedure 1 auction is guaranteed by the sampling based scheme. Therefore, to prove the truthfulness of TARCO is to prove the truthfulness of procedure 2 auction. For the SUE $r_i\in \mathcal{R}_w$, it can obtain the maximum utility $\max_{e_{k}\in \mathcal{E}}(B_i^{k}-A_k)$. If it bids $B_i^{k}=F_i(k)$, it can get the true utility as it bids. For SUE $r_i\notin \mathcal{R}_w$, it will lose the auction although it bids truthfully because $\max_{e_{k}\in \mathcal{E}}(B_i^{k}-A_k)<0$. If SUE $r_i$ bids untruthfully, there are two sub-cases. If $B_i^{k}<F_i(k)$, then the auction result will not change since SUE can only win the auction when $B_i^{k}>F_i(k)$. Even If $r_i$ wins the auction, its utility $U_i=F_i(k)-P_i=F_i(k)-B_{i}^{k}\le \max_{e_{k}\in \mathcal{E}}(B_i^{k}-B_i^{k})=0 $. Therefore, the SUE cannot improve its utility by trading untruthfully and this theorem holds.
\end{proof}
\end{theorem}

\begin{theorem}
	TARCO is individual rational and budget balance.
	\begin{proof}
	Firstly, we prove that the MUE $s_i^{j}$ is individual rational. Then we prove the SUE $r_i$ is also individual rational. According to Line $7$ of Algorithm \ref{alg:comptbgt}, we know that $v_i^{j}(k)>p_i^{j}(k)$ for $\forall s_i^{j}\in \mathcal{S}_w$. Then $u_i^{j}=v_i^{j}(k)-p_i^{j}(k)>0$ and we have proved the individual rationality of MUEs. To prove the rationality of SUEs, we only need to consider that the SUE $r_i$ wins in the procedure 2 auction because in other cases the utility of $r_i$ is zero. For $r_i$, $B_i^{k}\le F_i(k)$, and the winning SUE $r_i$ has to pay for the asks of the corresponding SCB $e_k$, \ie $P_i=A_k$ for $\forall r_i\in \mathcal{R}_w$ and $\forall e_k \in \mathcal{E}_w $. According to Line 3 of Algorithm \ref{algo:ph2}, for winning buyers and sellers in tier 2 Auction, we choose the maximum value of $B_i^{k}-A_k$. Therefore, $U_i=F_i(k)-P_i\ge B_i^{k}-A_k>0$. In TARCO, the auctioneer is always willing to help to coordinate the auction scheme, thus the utility of the auctioneer is always set to zero. Hence the proposed TARCO scheme is individual rational and budget balance.
		\end{proof}
\end{theorem}

\begin{theorem}
	{TARCO is computational efficient. The time complexity of phase I auction is $O(n_i \log n_i)$ and the time complexity of phase II auction is $O(MN)$.}
	\begin{proof}
For TARCO, in Algorithm \ref{alg:ph1}, the sorting process has a time complexity of $O(n_i \log n_i)$. The random splitting procedure takes $O(n)$ time. Algorithm \ref{alg:comptbgt} takes $O(n_i\log n_i)$ time. Therefore, the time complexity of Algorithm \ref{alg:ph1} is $O(n_i \log n_i)$. In Algorithm \ref{algo:ph2}, the outer `for' circulation takes time $O(N)$ and to acquire the winning SCB $e_k$, $k\in \mathcal{E}_w$ takes $O(M)$ time. Therefore, the time complexity of Algorithm \ref{algo:ph2} is $O(MN)$.

 Thus, TARCO has the property of computational efficiency.
		\end{proof}
\end{theorem}
\section{Improved Schemes VITA}\label{sec:improve}
In phase II of TARCO, the algorithm assigns SCB resources to the relay SUEs in a greedy manner, which may result in low utilities for both parties. Therefore, the SCBs may be reluctant to sell their computation resources and system performance may suffer. In this section, we propose a \textbf{V}CG based utility \textbf{i}mproved \textbf{t}ruthful \textbf{a}uction (VITA) scheme to improve the utilities of both SUEs and SCBs. In VITA, phase II of TARCO is replaced by a VCG based auction shown in Algorithm \ref{alg:vcg} while phase I of TARCO remains unchanged in the new scheme VITA. In VITA, to devise VCG based payment, we utilize a max weighted matching graph to determin winners. In this way, the total utility of SUEs and SCBs, which is expressed as $\sum_{k=1}^{M}(B_i^k-A_k)$ can be maximized. Different from the stage II of TASG presented in \cite{Lin2013Groupon} where a random maximum matching is adopted, we try to maximize the total utility of service prodivers in the two stages of TARCO and at the same time enrue the truthfulness of VITA. Since max weighted matching has a high computation complexity, in VITA, we utilise a 2-approximation ratio \cite{preis1999linear} of maximal weighted matching algorithm to determine winners. 

\begin{algorithm}[h]
\begin{algorithmic}[1]\caption{Phase II of VITA}\label{alg:vcg}
    \REQUIRE{$B_i^{k}$, for $\forall 1\le i \le N$ and $1\le k \le M$}
	\ENSURE{${\mathcal{R}}_w$, ${\mathcal{E}}_w$, $P_i$}
	\STATE{$\Lambda \leftarrow \emptyset$, $E^{*}\leftarrow \emptyset$ // $\Lambda$ is the edge set in the matching graph}
	\STATE{$P_i \leftarrow 0$}
	\STATE{Construct a bipartite graph $G=(\mathcal{R},\mathcal{E},\Lambda, w)$ and the weight of $
		\Lambda(r_i,e_k)=B_i^k - A_k$ if $B_i^k \ge A_k$ \\ /* $\mathcal{R}$ is the relay SUEs, $\mathcal{E}$ is set of SCBs, $\Lambda$ is the edge in the matching graph, $\forall (r_i,e_k)\in \Lambda$*/}
		\STATE{${E}^{*}\leftarrow \Lambda$ // Note that $w(E^{*})$ is the sum weight}
      \WHILE{$E^{*} \ne \emptyset$}
	   \STATE{Find an edge $(a,b)\in G$ with the highest weight}
	   \STATE{Add $(a,b)$ to the matching ${E}^{*}$ and delete all edges that incident to $a$ or $b$ in $G$}
	\ENDWHILE
	\FOR{each $(r_i,e_k)\in {E}^{*}$ }
	\STATE{$\mathcal{R}_w \leftarrow \mathcal{R}_w \cup \{r_i\},\ \mathcal{E}_w \leftarrow  \mathcal{E}_w \cup \{e_k\}$}
	\STATE{$\Lambda' \leftarrow \Lambda\backslash (r_i,e_k),\ \mathcal{R}'\leftarrow \mathcal{R} \backslash\{r_i\}$}
	\STATE{$G_{-i} \leftarrow (\mathcal{R}', \mathcal{E}, \Lambda', w)$}
	\STATE{ ${E}_{-i}^{*} \leftarrow \emptyset$}
	\WHILE{$\Lambda'\ne \emptyset$}
    \STATE{Find an edge $(c,d) \in G_{-i}$ with the highest weight}
    \STATE{Add $(c,d)$ to the matching ${E}_{-i}^{*}$ and delete all edges that incident to $c$ or $d$ in $G_{-i}$}
    \ENDWHILE
    \STATE{$P_i\leftarrow w({E}_{-i}^{*})-(w({E}^{*})-w(r_i,e_k))+A_k$}
    	\ENDFOR
    \STATE{Return ${\mathcal{R}}_w$, ${\mathcal{E}}_w$, $P_i$}
\end{algorithmic}
\end{algorithm} 

\begin{theorem}\label{theorem:2app}
	VITA satisfies the economic properties such as truthful, individual rational and computational efficient.
\begin{proof}
   Firstly, we prove that VITA is truthful. For each SUE relay node $r_i$, its bid is $B_i^{k}$, $\forall k\in [1,M]$. Its utility is calculated as $U_i=F_i(k)-P_i=F_i(k)-(w({E}_{-i}^{*})-(w({E}^{*})-w(r_i,e_k))+A_k)=w({E}^{*})-w(E_{-i}^{*})$. It should be noted that $w(E_{-i}^{*})$ is independent of SUE relay $r_i$, the maximum utility of $r_i$ can be obtained when $r_i$ bids $B_i^k=F_i(k)$, no $r_i$ can improve its utility if bidding $B_i^k\ne F_i(k)$. Therefore, VITA is truthful. 
   
Next, for any SUE $r_i$, its utility $U_i=F_i(k)-P_i=F_i(k)-(w({E}_{-i}^{*})-(w({E}^{*})-w(r_i,e_k))+A_k)=w({E}^{*})-w(E_{-i}^{*})\ge 0$, therefore, VITA is individual rational for the SUEs. 

Finally, based on the observation that, in VITA, phase I is just as same as that in TARCO, we only need to prove VITA is computational rational in phase II. The time complexity of the 2-approximation ratio algorithm of the maximum weighted matching is $O(Nlog(N+M))$ \cite{preis1999linear}, and the outer `for' circulation in Algorithm \ref{alg:vcg} takes time $O(N)$, thus the total time complexity is of Algorithm \ref{alg:vcg} is $O(N^2log(N+M))$.
\end{proof}
\end{theorem}
\section{Performance Evaluation}\label{sec:simulation}
In this section, we evaluate the performance of TARCO and make some comparisons between TARCO and random method. Meanwhile, we also derive the upper bound for TARCO since there is no existing algorithms to directly compare with. Then we investigate the performance of VITA with the method that use max weighted matching to determin winners. The simulation is executed on MATLAB simulator. 

\subsection{Simulation environment}
We consider a heterogeneous network structure shown in Fig. \ref{fig:network}, whose nodes are randomly distributed in a $100\times100$ area. The parameters for cooperative communications are adopted from \cite{chen2016two}. 

By default, we set $M=10$, $N=10$. We vary $n_i$ from $10$ to $100$ with an increment of $10$ for any $i\in \{1,2,\cdots,N\}$. We assume that MUE $s_i^{j}$'s budget $b_i^{j}(k)$ is randomly distributed in $(0,v_i^{j})$ , where $v_i^{j}$ is computed as described in Section \ref{subsec:sysmodel}. In phase II of TARCO, the ask price $A_k$ is randomly distributed in $(0,1]$. The results are averaged for 100 repetitions.

To the best of authors' knowledge, this is the first incentive scheme proposed for cooperative computation offloading in heterogeneous network and there are no existing auction schemes to compare with. Instead, in this paper, we compare TARCO with the following two schemes, \ie the upper bound and random schemes. To derive the upper bound, we choose the maximum $p_i^j(k)$ as $b_i^j(k)$ to replace that of Line $9$ in Algorithm \ref{alg:comptbgt}. Note that through this change, we do not change the economic properties of TARCO. For the random scheme, it replace the phase I of TARCO by fistly choosing a MUE $s_i^{t}$, if the valuation of another SU $s_i^{j}$ is larger than $s_i^{t}$, where $j\ne t$ and $j\in [1,n_i]$, SUE will allocate the computation resource to it. Then we examine the performance of utility enhanced sceme VITA with the above mentioned schemes as well as the \textbf{m}ax \textbf{w}eighted matching based winner \textbf{d}etermination scheme (MWD). 

\subsection{Simulation analysis}
\begin{figure}[ht]
	\centering
	\includegraphics[width=4.5in]{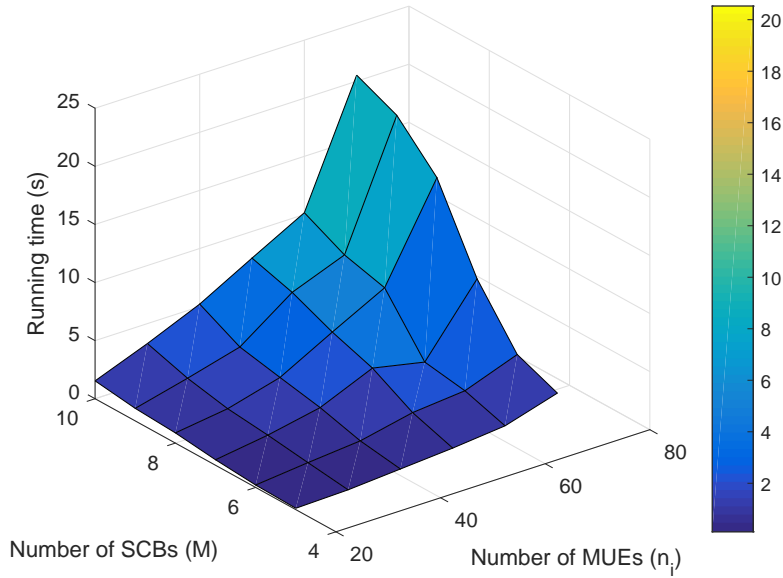}
	\caption{TARCO's running time with number of SUEs and number of SCBs (See digital or online version for color)}\label{fig:tarcort}
\end{figure}

Firstly, we investigate the running time of TARCO with different network settings shown in Fig. \ref{fig:tarcort}. In all simulations, the number of SUEs is set as same as that of SCBs' number, that is $N=M$. From Fig. \ref{fig:tarcort}, we observe that the running time is no more than $25$s when $n_i$ is lower than $80$. When $n_i$ is constant, the running time grows fast with the increasing of the SCB's number $M$ and meanwhile, when $M$ stays unchaged, the running time increases when $n_i$ becomes higher.

\begin{figure}
	\centering
	\includegraphics[width=4.5in]{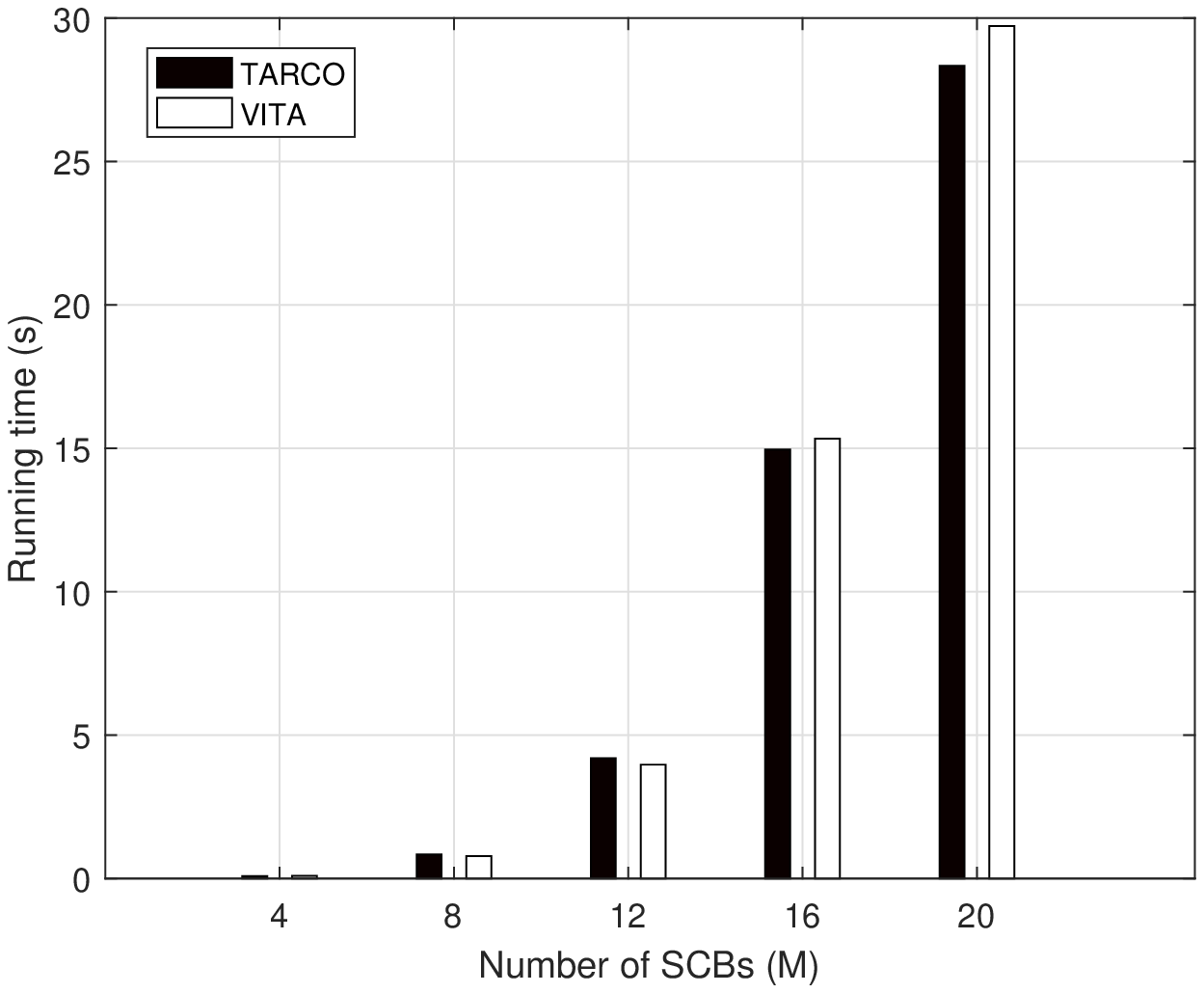}
	\caption{Running time comparisons between TARCO and VITA with number of SCBs }\label{fig:rtTVnumM}
\end{figure}

\begin{figure}
	\centering
	\includegraphics[width=4.5in]{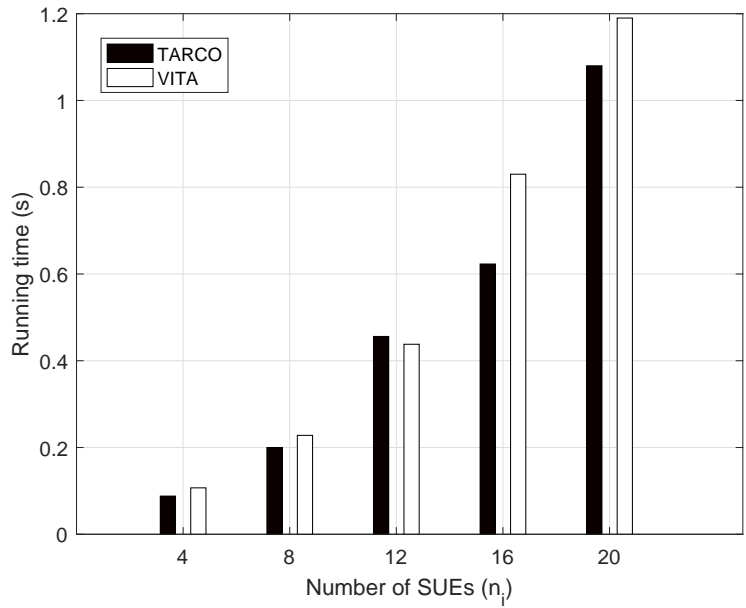}
	\caption{Running time comparisons between TARCO and VITA with number of SUEs }\label{fig:rtTVnumni}
\end{figure}



Next, we compare the running time performance between TARCO and VITA with different number of SCBs and SUEs respectively. The results are shown in Fig. \ref{fig:rtTVnumM} and Fig. \ref{fig:rtTVnumni}. We can see the running time performance of VITA is as similar as that of TARCO. In Fig. \ref{fig:rtTVnumM}, we set the number of SUEs as $n_i=20$ while we set the number of SCBs as $M=4$. It is obvious that when network size becomes larger, the running time of VITA is slightly higher than TARCO on average. That's because VITA tries to maximize the total utility which consumes much more time than TARCO.   

\begin{figure}[t]
	\centering
	\includegraphics[width=4.5in]{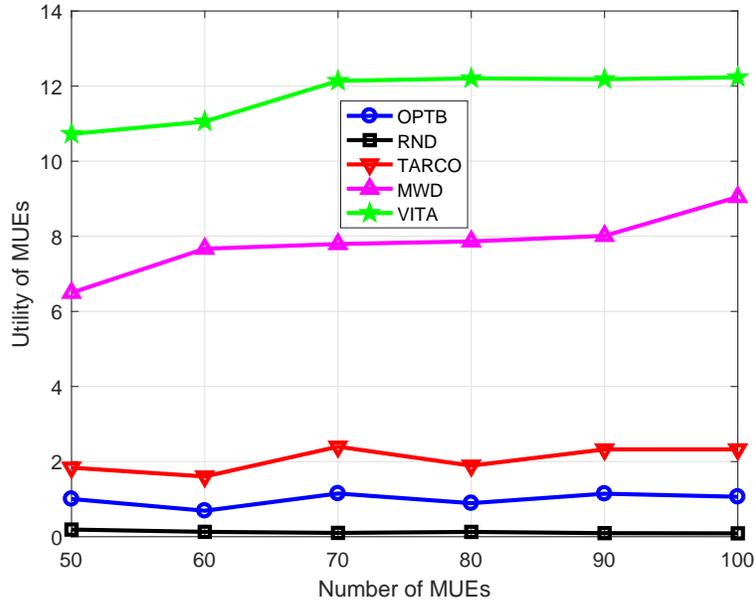}
	\caption{Average Utility of MUEs with the number of MUEs }\label{fig:UMUEsvsnumMUEs}
\end{figure}

\begin{figure}[t]
	\centering
	\includegraphics[width=4.5in]{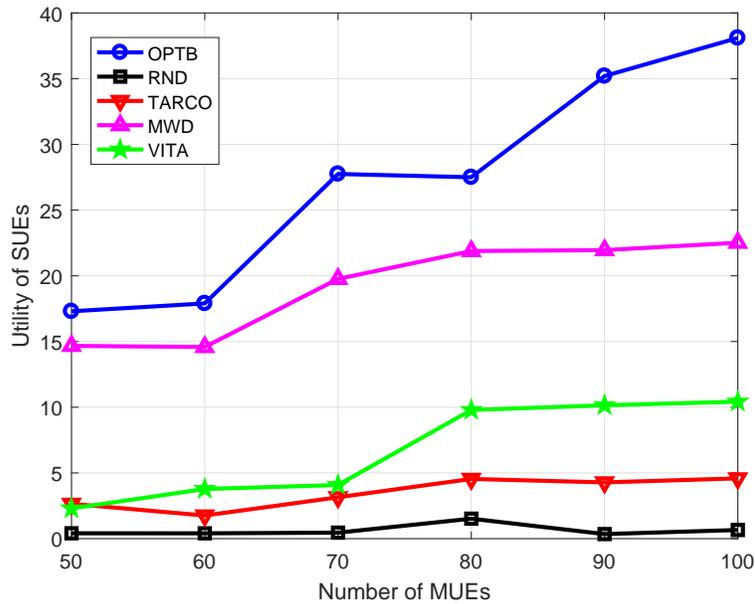}
	\caption{Average Utility of SUEs with the number of MUEs }\label{fig:USUEsvsnumMUEs}
\end{figure}

In the third experiment, we examine the utilities of network users for proposed schemes with $M=5$ and we vary $n_i$ from $50$ to $100$. As shown in Fig. \ref{fig:UMUEsvsnumMUEs}, we compre the average utility of MUEs with the increasing numnber of MUEs. As shown in Fig. \ref{fig:UMUEsvsnumMUEs}, with the increasing number of MUEs, the average utility of MUEs increases accordingly for MWD and VITA algorithms. For TARCO and OPTB, the average utility of MUEs varies around $2.0$ and $1.0$ respectively. Of all the algorithms, the random algorithm RND achieves the lowest utility in all situations. For MWD and VITA, the utility of MUEs grows fast when the number of MUEs is below $70$. That's because the competition for SUEs is not fierce since there are abundant resources. However, the utility gain decreases when the number of MUEs is above $70$ for MWD and VITA. That's because the CPU computation cycles obtained has reached to the limit and they cannot improve the utilities any longer. The utility of MUEs decreases with the increasing number of MUEs for the RND algorithm. That's because the RND algorithm tends to select inefficient MUEs. On average, OPTB achieves $46.90\%$ higher utility of MUEs than RND algorithm while TARCO is $104.90\%$ better than RND algorithm. What's more, VITA and MWD algorithms are the best two algorithms to enhance the utility of MUEs. VITA is $28.75\%$ better than TARCO and MWD outperforms TARCO for about $17.06\%$. That's because the two algorithms are based on max-weighted matching which greatly reduces the cost for MUEs.

Fig. \ref{fig:USUEsvsnumMUEs} shows the relationship between the average utility of SUEs and the number of MUEs. Generally, with the increasing number of MUEs, the average utility of SUEs grows accordingly. Different from Fig. \ref{fig:UMUEsvsnumMUEs}, the OPTB and MWD algorithms are among the best two algorithms. With the increasing number of MUEs, the utility gains of SUEs decrease for VITA, TARCO and RND algorithms when the number of MUEs is greater than $80$. For OPTB, it acquires the maximal $F_i(i)$ of all algorithms because $p_i^j(k)=b_i^j(k)$ for OPTB while for the other algorithms, $P_i^j(k)<b_i^j(k)$. Even though the utility gains of SUEs do not increase, the valuation of MUE $s_i^{j}$ is defined as (\ref{eq:vijkdef}), the SUE relay may select the MUEs with larger valuation and low capacity for direct communication, which result in the increase of utility of SUEs. On average, OPTB is about $317.65\%$ better than RND algorithm and TARCO achieves about $33.90\%$ utility gain than the RND algorithm on the utility of SUEs. Although VITA achieves almost the same utility of SUEs as TARCO, it outperforms TARCO by about $5.11\%$ whereas MWD outperforms TARCO by about $28.94\%$. 

\begin{figure}[t]
	\centering
	\includegraphics[width=4.5in]{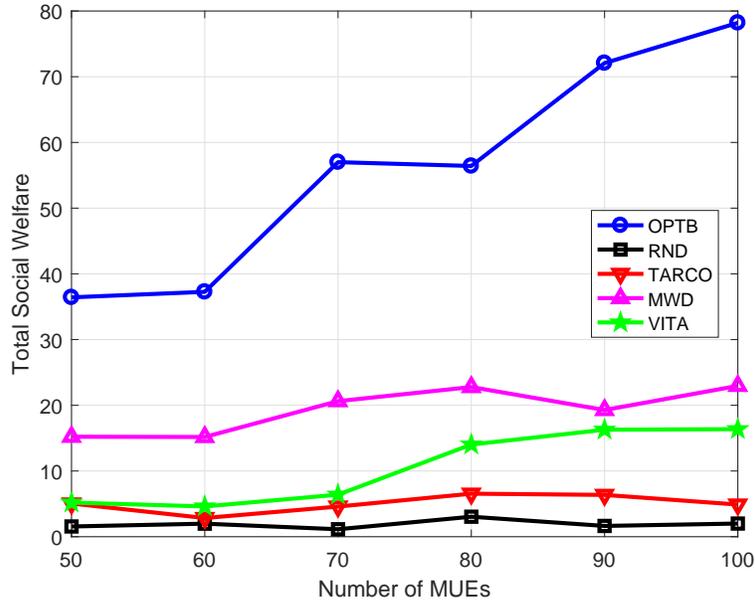}
	\caption{Social Welfare with the number of MUEs }\label{fig:TWvsnumMUEs}
\end{figure}

\begin{figure}[t]
	\centering
	\includegraphics[width=4.5in]{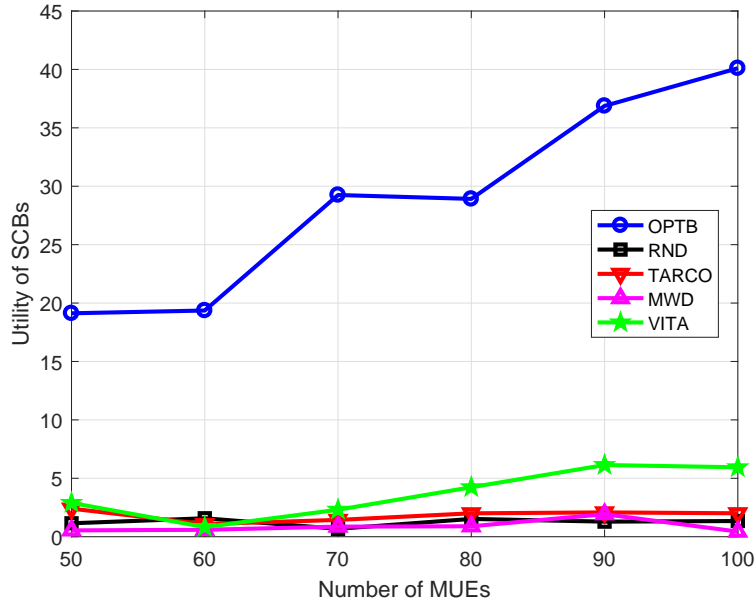}
	\caption{Average Utility of SCBs with the number of MUEs }\label{fig:USCBsvsnumMUEs}
\end{figure}

Fig. \ref{fig:TWvsnumMUEs} demonstrate almost the same trends as that of Fig. \ref{fig:USUEsvsnumMUEs}. Although OPTB achieves the best total social welfare, it cannot be achieved and it cannot fully guarantee the economic properties. Except for OPTB, MWD and VITA are among the best two algorithms that obtain the maximal utility. That's because the two algorithms try to maximize the sum utility of SCBs and SUEs, \ie
\begin{equation}
\sum_{i=1}^{N}\sum_{k=1}^{M}(F_i(k)-A_k),
\end{equation}
 based on max weighted matching. In conclusion, the OPTB algorithm achieves higher social utility than RND algorithm for about $188.04\%$ and TARCO is about $12.05\%$ better than RND. Compared with RND, MWD and VITA outperform RND by $63.14\%$ and $28.01\%$ respectively.

Finally, Fig. \ref{fig:USCBsvsnumMUEs} depicts the average utility of SCBs with the varying number of MUEs. It is obvious that the three algorithms, \ie TARCO, MWD and RND, produce similar average utility of SCBs. That's because under the simulation setting, the budget acquired by the SUE from MUEs is close to $A_k$ for the three algorithms. In gerneraly, OPTB can ensure a higher utility of SCBs, which is $143.67\%$ higher than RND algorithm whereas VITA is about $12.42\%$ better than RND. 

From the above experiments, we can see that VITA suits for the situation where all network participants' utility should be maximized at the same time. When the SCB is energy abundant and the cost can be neglected, TARCO, MWD and VITA demonstrate good qualities to ensure higher social welfare than the RND algorithm. In real situations, the SCB is a fixed network infrastructure, and is always full of computation resources as well as energy abundant. Therefore, TARCO, MWD and VITA can be deployed to motivate SUEs leasing resources to assist the computation offloading of MUEs.
\section{Conclusion}\label{sec:conclude}
In this paper, we have investigated the auction mechanism to jointly allocate the computation resources of SCBs and SUE relay nodes efectively in HetNet. We model the joint resource allocation problem as a two-stage auction where in the first stage, each remote MUE node submits its valuation and demand to bid SUE relay and the SCB. The SUE relay node collects bids from the social group within its range and conducts a virtual auction between MUEs and SUEs. Then the macro cell base station will perform a double auction between SUEs and SCBs to allocate the computation resources, \ie the CPU cycles. Finally, the winning SUE relays will execute the auction result of previous stage. Based on the proposed scheme, we have proposed three schemes TARCO, MWD and VITA. We prove that TARCO is truthful, individual rational, budget balance and computational efficient. We also prove that MWD and VITA schemes are truthful, individual rational and computational efficient. Extensive numerical analysis demonstrate that for social welfare, TARCO is about $12.05\%$ better than RND and Compared with RND, MWD and VITA outperform RND by $63.14\%$ and $28.01\%$ respectively. Note that in this work we assume SCBs have abundant resources and each SUE social group may only be served by one SCB and the auction is executed for one time. In the future, we would like to consider the auction with resource constraint SCBs and from a long time multi-round perspective. 


\bibliographystyle{IEEEtran}
\bibliography{refs}

\begin{thebibliography}{10}
\providecommand{\url}[1]{#1}
\csname url@samestyle\endcsname
\providecommand{\newblock}{\relax}
\providecommand{\bibinfo}[2]{#2}
\providecommand{\BIBentrySTDinterwordspacing}{\spaceskip=0pt\relax}
\providecommand{\BIBentryALTinterwordstretchfactor}{4}
\providecommand{\BIBentryALTinterwordspacing}{\spaceskip=\fontdimen2\font plus
\BIBentryALTinterwordstretchfactor\fontdimen3\font minus
  \fontdimen4\font\relax}
\providecommand{\BIBforeignlanguage}[2]{{%
\expandafter\ifx\csname l@#1\endcsname\relax
\typeout{** WARNING: IEEEtran.bst: No hyphenation pattern has been}%
\typeout{** loaded for the language `#1'. Using the pattern for}%
\typeout{** the default language instead.}%
\else
\language=\csname l@#1\endcsname
\fi
#2}}
\providecommand{\BIBdecl}{\relax}
\BIBdecl

\bibitem{indexcisco}
Cisco, ``Cisco visual networking index: Global mobile data traffic forecast
  update, 2015--2020 white paper,''
  \url{http://www.cisco.com/c/en/us/solutions/collateral/service-provider/visual-networking-index-vni/mobile-white-paper-c11-520862.html},
  2016, [Online; accessed 11-Nov-2016].

\bibitem{zhu2014pricing}
K.~Zhu, E.~Hossain, and D.~Niyato, ``Pricing, spectrum sharing, and service
  selection in two-tier small cell networks: A hierarchical dynamic game
  approach,'' \emph{IEEE Transactions on Mobile Computing}, vol.~13, no.~8, pp.
  1843--1856, 2014.

\bibitem{jutruthful}
P.~Ju and W.~Song, ``Truthful budget constrained auction for device-to-device
  relaying packet allocation,'' \emph{Wireless Networks}, vol.~22, no.~7, pp.
  2173--2188.

\bibitem{xing2010investigation}
H.~Xing and S.~Hakola, ``The investigation of power control schemes for a
  device-to-device communication integrated into ofdma cellular system,'' in
  \emph{21st Annual IEEE International Symposium on Personal, Indoor and Mobile
  Radio Communications}.\hskip 1em plus 0.5em minus 0.4em\relax IEEE, 2010, pp.
  1775--1780.

\bibitem{yin2015pricing}
R.~Yin, G.~Yu, H.~Zhang, Z.~Zhang, and G.~Y. Li, ``Pricing-based interference
  coordination for d2d communications in cellular networks,'' \emph{IEEE
  Transactions on Wireless Communications}, vol.~14, no.~3, pp. 1519--1532,
  2015.

\bibitem{song2014game}
L.~Song, D.~Niyato, Z.~Han, and E.~Hossain, ``Game-theoretic resource
  allocation methods for device-to-device communication,'' \emph{IEEE Wireless
  Communications}, vol.~21, no.~3, pp. 136--144, 2014.

\bibitem{malandrino2015interference}
F.~Malandrino, Z.~Limani, C.~Casetti, and C.-F. Chiasserini,
  ``Interference-aware downlink and uplink resource allocation in hetnets with
  d2d support,'' \emph{IEEE Transactions on Wireless Communications}, vol.~14,
  no.~5, pp. 2729--2741, 2015.

\bibitem{shi2008optimal}
Y.~Shi, S.~Sharma, Y.~T. Hou, and S.~Kompella, ``Optimal relay assignment for
  cooperative communications,'' in \emph{Proceedings of the 9th ACM
  international symposium on Mobile ad hoc networking and computing}.\hskip 1em
  plus 0.5em minus 0.4em\relax ACM, 2008, pp. 3--12.

\bibitem{ur2015analysis}
A.~ur~Rahman and S.~A. Hassan, ``Analysis of composite fading in a single cell
  downlink cooperative heterogeneous networks,'' in \emph{the 81st Vehicular
  Technology Conference (VTC Spring)}.\hskip 1em plus 0.5em minus 0.4em\relax
  IEEE, 2015, pp. 1--5.

\bibitem{xiao2015optimal}
S.~Xiao, D.~Feng, Y.~Yuan-Wu, G.~Y. Li, W.~Guo, and S.~Li, ``Optimal mobile
  association in device-to-device-enabled heterogeneous networks,'' in
  \emph{Vehicular Technology Conference (VTC Fall), 2015 IEEE 82nd}.\hskip 1em
  plus 0.5em minus 0.4em\relax IEEE, 2015, pp. 1--5.

\bibitem{xiao2016energy}
S.~Xiao, X.~Zhou, D.~Feng, Y.~Yuan-Wu, G.~Y. Li, and W.~Guo, ``Energy-efficient
  mobile association in heterogeneous networks with device-to-device
  communications,'' \emph{IEEE Transactions on Wireless Communications},
  vol.~15, no.~8, pp. 5260--5271, 2016.

\bibitem{chen2015load}
Z.~Chen, H.~Zhao, Y.~Cao, and T.~Jiang, ``Load balancing for d2d-based relay
  communications in heterogeneous network,'' in \emph{Modeling and Optimization
  in Mobile, Ad Hoc, and Wireless Networks (WiOpt), 2015 13th International
  Symposium on}.\hskip 1em plus 0.5em minus 0.4em\relax IEEE, 2015, pp. 23--29.

\bibitem{hwang2015ue}
D.~Hwang, D.~I. Kim, S.~K. Choi, and T.-J. Lee, ``Ue relaying cooperation over
  d2d uplink in heterogeneous cellular networks,'' \emph{IEEE Transactions on
  Communications}, vol.~63, no.~12, pp. 4784--4796, 2015.

\bibitem{jiang2015rally}
J.~Jiang, Y.~Zhu, B.~Li, and B.~Li, ``Rally: Device-to-device content sharing
  in lte networks as a game,'' in \emph{Mobile Ad Hoc and Sensor Systems
  (MASS), 2015 IEEE 12th International Conference on}.\hskip 1em plus 0.5em
  minus 0.4em\relax IEEE, 2015, pp. 10--18.

\bibitem{zhang2015contract}
Y.~Zhang, L.~Song, W.~Saad, Z.~Dawy, and Z.~Han, ``Contract-based incentive
  mechanisms for device-to-device communications in cellular networks,''
  \emph{IEEE Journal on Selected Areas in Communications}, vol.~33, no.~10, pp.
  2144--2155, 2015.

\bibitem{chen2015efficient}
X.~Chen, L.~Jiao, W.~Li, and X.~Fu, ``Efficient multi-user computation
  offloading for mobile-edge cloud computing,'' \emph{IEEE/ACM Trans. on
  Networking}, 2015, doi:10.1109/TNET.2015.2487344.

\bibitem{sardellitti2015joint}
S.~Sardellitti, G.~Scutari, and S.~Barbarossa, ``Joint optimization of radio
  and computational resources for multicell mobile-edge computing,'' \emph{IEEE
  Transactions on Signal and Information Processing over Networks}, vol.~1,
  no.~2, pp. 89--103, 2015.

\bibitem{semiari2015context}
O.~Semiari, W.~Saad, S.~Valentin, M.~Bennis, and H.~V. Poor, ``Context-aware
  small cell networks: How social metrics improve wireless resource
  allocation,'' \emph{IEEE Transactions on Wireless Communications}, vol.~14,
  no.~11, pp. 5927--5940, 2015.

\bibitem{cao2016share}
Y.~Cao, C.~Long, T.~Jiang, and S.~Mao, ``Share communication and computation
  resources on mobile devices: a social awareness perspective,'' \emph{IEEE
  Wireless Communications}, vol.~23, no.~4, pp. 52--59, 2016.

\bibitem{Lin2013Groupon}
P.~Lin, X.~Feng, Q.~Zhang, and M.~Hamdi, ``Groupon in the air: A three-stage
  auction framework for spectrum group-buying,'' \emph{Proceedings - IEEE
  INFOCOM}, vol.~12, no.~11, pp. 2013--2021, 2013.

\bibitem{chen2015spectrum}
L.~Chen, L.~Huang, Z.~Sun, H.~Xu, and H.~Guo, ``Spectrum combinatorial double
  auction for cognitive radio network with ubiquitous network resource
  providers,'' \emph{IET Communications}, vol.~9, no.~17, pp. 2085--2094, 2015.

\bibitem{jo2015device}
M.~Jo, T.~Maksymyuk, B.~Strykhalyuk, and C.-H. Cho, ``Device-to-device-based
  heterogeneous radio access network architecture for mobile cloud computing,''
  \emph{IEEE Wireless Communications}, vol.~22, no.~3, pp. 50--58, 2015.

\bibitem{yang2014truthful}
D.~Yang, G.~Xue, and X.~Zhang, ``Truthful group buying-based spectrum auction
  design for cognitive radio networks,'' in \emph{2014 IEEE International
  Conference on Communications (ICC)}.\hskip 1em plus 0.5em minus 0.4em\relax
  IEEE, 2014, pp. 2295--2300.

\bibitem{liu2015device}
J.~Liu and N.~Kato, ``Device-to-device communication overlaying two-hop
  multi-channel uplink cellular networks,'' in \emph{Proceedings of the 16th
  ACM International Symposium on Mobile Ad Hoc Networking and Computing
  (MobiHoc)}.\hskip 1em plus 0.5em minus 0.4em\relax ACM, 2015, pp. 307--316.

\bibitem{cao2016cellular}
W.~Cao, G.~Feng, S.~Qin, and M.~Yan, ``Cellular offloading in heterogeneous
  mobile networks with d2d communication assistance,'' \emph{IEEE Transactions
  on Vehicular Technology}, 2016.

\bibitem{kawamoto2014efficient}
Y.~Kawamoto, J.~Liu, H.~Nishiyama, and N.~Kato, ``An efficient traffic
  detouring method by using device-to-device communication technologies in
  heterogeneous network,'' in \emph{2014 IEEE Wireless Communications and
  Networking Conference (WCNC)}.\hskip 1em plus 0.5em minus 0.4em\relax IEEE,
  2014, pp. 2162--2167.

\bibitem{hua2013truthful}
S.~Hua, X.~Zhuo, and S.~S. Panwar, ``A truthful auction based incentive
  framework for femtocell access,'' in \emph{2013 IEEE Wireless Communications
  and Networking Conference (WCNC)}.\hskip 1em plus 0.5em minus 0.4em\relax
  IEEE, 2013, pp. 2271--2276.

\bibitem{leanh2015joint}
T.~LeAnh, N.~H. Tran, S.~A. Kazmi, T.~Z. Oo, and C.~S. Hong, ``Joint pricing
  and power allocation for uplink macrocell and femtocell cooperation,'' in
  \emph{2015 International Conference on Information Networking (ICOIN)}.\hskip
  1em plus 0.5em minus 0.4em\relax IEEE, 2015, pp. 171--176.

\bibitem{ho2016coordinated}
T.~M. Ho, N.~H. Tran, L.~B. Le, W.~Saad, S.~A. Kazmi, and C.~S. Hong,
  ``Coordinated resource partitioning and data offloading in wireless
  heterogeneous networks,'' \emph{IEEE Communications Letters}, vol.~20, no.~5,
  pp. 974--977, 2016.

\bibitem{wang2015truthful}
X.~Wang, L.~Huang, H.~Xu, and H.~Huang, ``Truthful auction for resource
  allocation in cooperative cognitive radio networks,'' in \emph{2015 24th
  International Conference on Computer Communication and Networks
  (ICCCN)}.\hskip 1em plus 0.5em minus 0.4em\relax IEEE, 2015, pp. 1--8.

\bibitem{jin2016auction}
A.-L. Jin, W.~Song, and W.~Zhuang, ``Auction-based resource allocation for
  sharing cloudlets in mobile cloud computing,'' \emph{IEEE Transactions on
  Emerging Topics in Computing}, pp. 1--12, 2016.

\bibitem{liu2016energy}
H.~Liu, H.~Zhang, J.~Cheng, and V.~C. Leung, ``Energy efficient power
  allocation and backhaul design in heterogeneous small cell networks,'' in
  \emph{Communications (ICC), 2016 IEEE International Conference on}.\hskip 1em
  plus 0.5em minus 0.4em\relax IEEE, 2016, pp. 1--5.

\bibitem{khawer2016usicic}
M.~R. Khawer, J.~Tang, and F.~Han, ``usicic—a proactive small cell
  interference mitigation strategy for improving spectral efficiency of lte
  networks in the unlicensed spectrum,'' \emph{IEEE Transactions on Wireless
  Communications}, vol.~15, no.~3, pp. 2303--2311, 2016.

\bibitem{fiat2002competitive}
A.~Fiat, A.~V. Goldberg, J.~D. Hartline, and A.~R. Karlin, ``Competitive
  generalized auctions,'' in \emph{Proceedings of the thiry-fourth annual ACM
  symposium on Theory of computing}.\hskip 1em plus 0.5em minus 0.4em\relax
  ACM, 2002, pp. 72--81.

\bibitem{goldberg2001competitive}
A.~V. Goldberg, J.~D. Hartline, and A.~Wright, ``Competitive auctions and
  digital goods,'' in \emph{Proceedings of the twelfth annual ACM-SIAM
  symposium on Discrete algorithms}.\hskip 1em plus 0.5em minus 0.4em\relax
  Society for Industrial and Applied Mathematics, 2001, pp. 735--744.

\bibitem{yang2011truthful}
D.~Yang, X.~Fang, and G.~Xue, ``Truthful auction for cooperative
  communications,'' in \emph{Proceedings of the Twelfth ACM International
  Symposium on Mobile Ad Hoc Networking and Computing}.\hskip 1em plus 0.5em
  minus 0.4em\relax ACM, 2011, pp. 1--9.

\bibitem{preis1999linear}
R.~Preis, ``Linear time 1/2-approximation algorithm for maximum weighted
  matching in general graphs,'' in \emph{Annual Symposium on Theoretical
  Aspects of Computer Science}.\hskip 1em plus 0.5em minus 0.4em\relax
  Springer, 1999, pp. 259--269.

\bibitem{chen2016two}
L.~Chen, J.~Wu, H.-N. Dai, and X.~Huang, ``Two-stage game for joint bandwidth
  and multiple homing relay allocation in cooperative d2d networks,'' in
  \emph{Proceedings of the 18th IEEE International Conference on High
  Performance Computing and Communications (HPCC)}.\hskip 1em plus 0.5em minus
  0.4em\relax IEEE, 2016, pp. 554--561.

\end{thebibliography}

\end{spacing}
\end{document}